\documentclass[12pt,a4paper,english,superscriptaddress,aps]{revtex4}
\usepackage{amssymb}
\usepackage{amsmath} 
\usepackage{slashed}
\usepackage{graphicx}
\usepackage{babel}
\usepackage{hyperref}
\usepackage{color}
\usepackage{comment}
\makeatletter\AtBeginDocument{\let\@elt\relax}\makeatother

\begin{document}

\title{Ultra-Planckian quark and gluon scattering in agravity}

\author{I. F. Cunha}
\email{ivana.cunha@icen.ufpa.br} 
\affiliation{Faculdade de F\'{i}sica, Universidade Federal do Par\'{a}, 66075-110, Bel\'{e}m, Par\'a, Brazil}

\author{A. C. Lehum}
\email[]{lehum@ufpa.br}
\affiliation{Faculdade de F\'{i}sica, Universidade Federal do Par\'{a}, 66075-110, Bel\'{e}m, Par\'a, Brazil}

\begin{abstract}
We investigate tree-level scattering processes involving quarks ($q$) and gluons ($g$) mediated by graviton exchange in the framework of agravity, a dimensionless and renormalizable theory of quadratic quantum gravity. Focusing on the ultra-Planckian regime, characterized by the Mandelstam variable $s = (p_1 + p_2)^2$, which corresponds to the total energy squared in the center-of-momentum frame, being much larger than any particle mass scale, we compute the squared amplitudes and analyze the differential cross sections for the processes $gg \to gg$, $gg \to q\bar{q}$, $gq \to gq$, and $qq \to qq$. We demonstrate that all amplitudes scale as $1/s$ at high energies, in agreement with expectations for a UV-complete theory of gravity. In addition, we explore the issue of unitarity in the presence of higher-derivative ghost modes by analyzing the positivity properties of the squared amplitudes. While IR divergences appear in the forward scattering of massless particles, we show that these are regularized by finite quark masses. Our findings support the viability of agravity as a perturbatively unitary and UV-complete extension of general relativity, capable of consistently describing gravitational interactions among elementary matter fields at trans-Planckian energies.
\end{abstract}

\maketitle

\section{Introduction}

It is well established that the quantization of Einstein's General Relativity (GR) around flat spacetime leads to a nonrenormalizable quantum field theory \cite{tHooft:1974toh,Deser:1974zzd,Deser:1974cy}. Among the various approaches to formulating a quantum theory of gravity, one possibility is to treat gravity as an effective field theory valid below the Planck scale, where quantum corrections can be systematically organized \cite{Donoghue:1994dn,Burgess:2003jk,Donoghue:2017pgk}. An alternative route is to seek a renormalizable UV completion that reduces to the Einstein–Hilbert action at low energies.

In this context, quadratic quantum gravity (QQG) has been extensively studied as a promising candidate for a renormalizable theory of quantum gravity \cite{Stelle:1976gc,Tomboulis:1977jk,Odintsov:1991nd,Salvio:2014soa,Salvio:2017qkx,Einhorn:2014gfa,Buchbinder:2017lnd,Donoghue:2021cza}. Although such theories are renormalizable, they typically suffer from issues related to the presence of ghostlike or tachyonic modes. Recent investigations \cite{Salvio:2014soa,Alvarez-Gaume:2015rwa,Holdom:2015kbf,Anselmi:2018ibi,Salvio:2018crh,Donoghue:2021cza,Buccio:2024hys} suggest that while ghost states might be controllable under certain prescriptions, the presence of tachyonic instabilities is generally regarded as a fundamental obstacle.

In Ref.~\cite{Salvio:2014soa}, the authors propose a model known as \emph{agravity}, which explores the possibility of constructing a gravitational theory devoid of any fundamental scale. This is achieved by including only quadratic curvature invariants in the pure gravity sector \cite{Aoki:2021skm,Aoki:2024jhr,Alvarez-Luna:2022hka}, leading to a renormalizable theory of quantum gravity characterized by a graviton kinetic term containing four derivatives. As a consequence, the graviton propagator behaves as $1/p^4$ in momentum space \cite{Buoninfante:2023ryt}. Within this framework, the authors further discuss the intriguing possibility that the Planck scale may emerge dynamically at the quantum level via dimensional transmutation~\cite{Gialamas:2020snr,Salvio:2020axm}.

Although agravity is free from tachyonic instabilities, it is not, in principle, exempt from the presence of ghostlike states. It is worth emphasizing that the issue of unitarity in QQG has been examined in the recent literature. In Ref.~\cite{Salvio:2019wcp}, it was shown that when properly quantized, QQG can remain unitary despite the presence of higher-derivative terms. This analysis was further extended in Ref.~\cite{Salvio:2024joi}, where the proof of unitarity was generalized to the nonperturbative and background-independent regime, reinforcing the consistency of these theories beyond perturbation theory. In addition, IR properties of agravity were studied in Ref.~\cite{Salvio:2018kwh}, where the authors identified an IR enhancement in tree-level scattering amplitudes involving scalar, fermionic, and gauge fields.

While these results strongly support the theoretical consistency of agravity at both the perturbative and nonperturbative levels, it is also instructive to assess how such consistency manifests in explicit scattering processes involving matter and gauge fields. Such amplitude-based studies offer an alternative window into the interplay between higher-derivative dynamics and unitarity, providing valuable insight into the physical viability of agravity when applied to high-energy processes involving gauge and matter fields. 

Progress in this direction has been made by Holdom in Refs.~\cite{Holdom:2021oii,Holdom:2021hlo}. In Ref.~\cite{Holdom:2021oii}, the author demonstrates that QQG leads to a well behaved differential cross section for photon-photon scattering across all energy scales. The study also highlights how unitarity may be preserved even in the absence of positivity, with significant implications for the structure of scattering amplitudes and cross sections. In a subsequent work, Ref.~\cite{Holdom:2021hlo}, it is shown that high-energy scattering cross sections in QQG exhibit a $1/E^2$ scaling due to remarkable cancellations among the contributions of different graviton modes. This behavior provides further support for the idea that QQG may constitute a viable UV-complete theory of quantum gravity, capable of maintaining unitarity without relying on the positivity of individual amplitudes.

Following the line of investigation initiated by Holdom, the study of quark and gluon scattering processes within the framework of UV-complete theories of gravity, such as QQG, can provide important insights into the consistency and universality of these models. Since gravity couples universally to the energy-momentum tensor, it is crucial to examine whether the unitarity-preserving cancellations observed in the photon-photon scattering also occur in interactions involving non-Abelian gauge fields. Quarks and gluons, as the fundamental constituents of QCD, introduce nontrivial color structures and gauge dynamics, rendering their scattering processes a natural way for exploring the interplay between gauge and gravitational sectors.

The main objective of the present work is to compute graviton-mediated scattering amplitudes for quarks and gluons within the framework of QQG, specifically in the ultra-Planckian regime. To this end, we couple massless QCD fields to agravity, a dimensionless and renormalizable model of QQG, and analyze the positivity of the squared scattering amplitudes, as well as the behavior of the corresponding differential cross sections in the deep UV.

The structure of the paper is as follows. In Sec.~\ref{sec02}, we present the action of the model and derive the corresponding field propagators. Sec.~\ref{sec03} is devoted to the computation of the scattering amplitudes mediated by agravity, with emphasis on their behavior in the deep UV regime. Final discussions and concluding remarks are provided in Sec.~\ref{summary}.

Throughout this paper, we use natural units $c=\hbar=1$.

\section{The QCD-agravity Lagrangian}\label{sec02}

We begin with the classical action describing a non-Abelian gauge theory coupled to fermions interacting with dimensionless gravity within the framework of agravity:
\begin{eqnarray}\label{eq01}
\mathcal{S} &=& \int d^4x\, \sqrt{-g} \Bigg\{ \frac{R^2}{6f_0^2}
+\frac{1}{f_2^2}\left(\frac{1}{3}R^2 - R^{\mu\nu}R_{\mu\nu}\right)
- \frac{1}{4} g^{\mu\alpha}g^{\nu\beta} G_{\mu\nu}^a G_{\alpha\beta}^a \nonumber\\
&& + i\, \bar{\psi}_f \left(\nabla_{\mu} - ig A_\mu^a t^a \right) \gamma^\mu \psi_f \Bigg\},
\end{eqnarray}
\noindent
where $R$ and $R_{\mu\nu}$ denote the Ricci scalar and Ricci tensor, respectively, and $\nabla_\mu$ is the covariant derivative. The index $f = 1, 2, \dots, N_f$ labels the fermion flavors. The non-Abelian field strength is given by $G^a_{\mu\nu} = \partial_\mu A_\nu^a - \partial_\nu A_\mu^a + g f^{abc} A_\mu^b A_\nu^c$, with $t^a$ representing the $SU(N)$ generators and $f^{abc}$ the corresponding structure constants.

The Dirac matrices are defined in terms of the vierbein by $\gamma^\mu = \gamma^\alpha e^\mu_\alpha$, where the spacetime metric satisfies $g_{\mu\nu} = e_\mu^\alpha e_\nu^\beta \eta_{\alpha\beta}$. The spinor covariant derivatives are defined as $\overrightarrow{\nabla}_\mu \psi = (\partial_\mu + i\omega_\mu)\psi$ and $\bar{\psi} \overleftarrow{\nabla}_\mu = \partial_\mu \bar{\psi} - i \bar{\psi} \omega_\mu$, where the spin connection is
\begin{eqnarray}
\omega_\mu = \frac{1}{4} \sigma^{\alpha\beta} \left[ e^\nu_\alpha (\partial_\mu e_{\beta\nu} - \partial_\nu e_{\beta\mu}) 
+ \frac{1}{2} e^\rho_\alpha e^\sigma_\beta (\partial_\sigma e_{\gamma\rho} - \partial_\rho e_{\gamma\sigma}) e^\gamma_\mu 
- (\alpha \leftrightarrow \beta) \right],
\end{eqnarray}
with $\sigma^{\alpha\beta} = \frac{i}{2}[\gamma^\alpha, \gamma^\beta]$. Throughout, Greek indices from the middle of the alphabet ($\mu, \nu, \dots$) refer to general spacetime coordinates, while those from the beginning of the alphabet ($\alpha, \beta, \dots$) denote locally inertial (tangent space) coordinates.

In order to study the model, we expand the spacetime metric around the flat Minkowski background as
\begin{equation}\label{metric} 	
g_{\mu\nu} = \eta_{\mu\nu} + h_{\mu\nu} \quad \text{(exact)}, \qquad g^{\mu\nu} = \eta^{\mu\nu} - h^{\mu\nu} + \cdots,
\end{equation}
where spacetime indices (Greek letters) are raised and lowered using the flat Minkowski metric $\eta_{\mu\nu} = \mathrm{diag}(+,-,-,-)$. Since our goal is to compute the tree-level scattering amplitudes involving quarks and gluons, we restrict the analysis to the one-graviton exchange approximation, and therefore retain only the linear terms in the gravitational field \( h_{\mu\nu} \).

Moreover, we do not include Faddeev–Popov ghost contributions, as they are relevant only for loop-level calculations. However, to define the graviton propagator, we introduce the gravitational gauge-fixing term
\begin{equation}
\mathcal{L}_{\text{GF},h} = -\frac{1}{2\zeta_g} \partial^\nu \left(h_{\mu\nu} - \frac{1}{2} \eta_{\mu\nu} h \right) \partial^\alpha \left(h^{\mu}_{\ \alpha} - \frac{1}{2} \eta^{\mu\alpha} h \right),
\end{equation}
and, analogously, the gauge-fixing term for the gluon field is given by
\begin{equation}
\mathcal{L}_{\text{GF},A} = -\frac{1}{2\xi} \left(\partial_\mu A^\mu_a\right)^2,
\end{equation}
where $\zeta_g$  and $\xi$ are the gravitational and Yang–Mills gauge-fixing parameters, respectively. The explicit forms of the linearized Lagrangians describing gravity coupled to non-Abelian gauge fields can be found in Refs.~\cite{Choi:1994ax,Souza:2022ovu, Souza:2023wzv, Gomes:2024onm}.

The quadratic part of the action yields the following propagators:
\begin{eqnarray}
\Delta_{\mu\nu}^{ab}(p) &=& -\frac{i}{p^2} \left[ T^{\mu\nu} + \xi\, L^{\mu\nu} \right] \delta^{ab}, \nonumber\\
\Delta_{\mu\nu\rho\sigma}(p) &=& \frac{i}{p^4} \left[ -2f_2^2\, P^{(2)}_{\mu\nu\rho\sigma} + f_0^2\, P^{(0)}_{\mu\nu\rho\sigma} 
+ 2\zeta_g \left( P^{(1)}_{\mu\nu\rho\sigma} + \frac{1}{2} P^{(0w)}_{\mu\nu\rho\sigma} \right) \right],
\end{eqnarray}
\noindent where $\Delta_{\mu\nu}^{ab}(p)$ and $\Delta_{\mu\nu\rho\sigma}(p)$ denote the gluon and graviton propagators, respectively. The tensor structures appearing in the propagators are defined in terms of the transverse and longitudinal projectors:
\begin{eqnarray}
T_{\mu\nu} &=& \eta_{\mu\nu} - \frac{p_\mu p_\nu}{p^2}, \nonumber\\
L_{\mu\nu} &=& \frac{p_\mu p_\nu}{p^2}.
\end{eqnarray}

The spin projection operators in the symmetric two-index tensor space are given by
\begin{eqnarray}
P^{(2)}_{\mu\nu\rho\sigma} &=& \frac{1}{2} T_{\mu\rho} T_{\nu\sigma} + \frac{1}{2} T_{\mu\sigma} T_{\nu\rho} - \frac{1}{D - 1} T_{\mu\nu} T_{\rho\sigma}, \nonumber\\
P^{(1)}_{\mu\nu\rho\sigma} &=& \frac{1}{2} \left( T_{\mu\rho} L_{\nu\sigma} + T_{\mu\sigma} L_{\nu\rho} + L_{\mu\rho} T_{\nu\sigma} + L_{\mu\sigma} T_{\nu\rho} \right), \nonumber\\
P^{(0)}_{\mu\nu\rho\sigma} &=& \frac{1}{D - 1} T_{\mu\nu} T_{\rho\sigma}, \nonumber\\
P^{(0w)}_{\mu\nu\rho\sigma} &=& L_{\mu\nu} L_{\rho\sigma}.
\end{eqnarray}

\noindent
Here, $P^{(2)}_{\mu\nu\rho\sigma}$, $P^{(1)}_{\mu\nu\rho\sigma}$, and $P^{(0)}_{\mu\nu\rho\sigma}$ project onto the spin-2, spin-1, and spin-0 components of the graviton field, respectively, while $P^{(0w)}_{\mu\nu\rho\sigma}$ accounts for the longitudinal spin-0 contribution. The projectors satisfy the standard completeness relations in the space of symmetric rank-2 tensors.

With all the necessary ingredients in place, we are now prepared to compute the scattering amplitudes for quarks and gluons mediated by a graviton within the framework of dimensionless quadratic gravity. This analysis will be carried out in the following section.

\section{Agravity-Mediated Scattering Processes}\label{sec03}

In this section, we compute the scattering processes mediated by agravity, specifically focusing on gluon–gluon scattering into two gluons, gluon–gluon scattering into a quark–antiquark pair, gluon–quark scattering, and quark–quark scattering. The relevant energy scale for agravity-mediated interactions lies beyond the Planck scale, $E > M_P$. For several of these processes, we have performed the calculations using an arbitrary choice of the gravitational gauge-fixing parameter $\zeta_g$, explicitly verifying that the final amplitudes are independent of $\zeta_g$, as required by gauge invariance. In the case of gluon–gluon to gluon–gluon scattering, however, we fixed $\zeta_g = 0$ in order to reduce the computational complexity. This choice does not entail any loss of generality, as the physical S-matrix is expected to be gauge independent. We begin by analyzing the gluon–gluon scattering process that results in two final-state gluons.

\subsection*{Gluon-gluon scattering to two gluons}

\begin{figure}[h!]
	\includegraphics[angle=0 ,width=13.5cm]{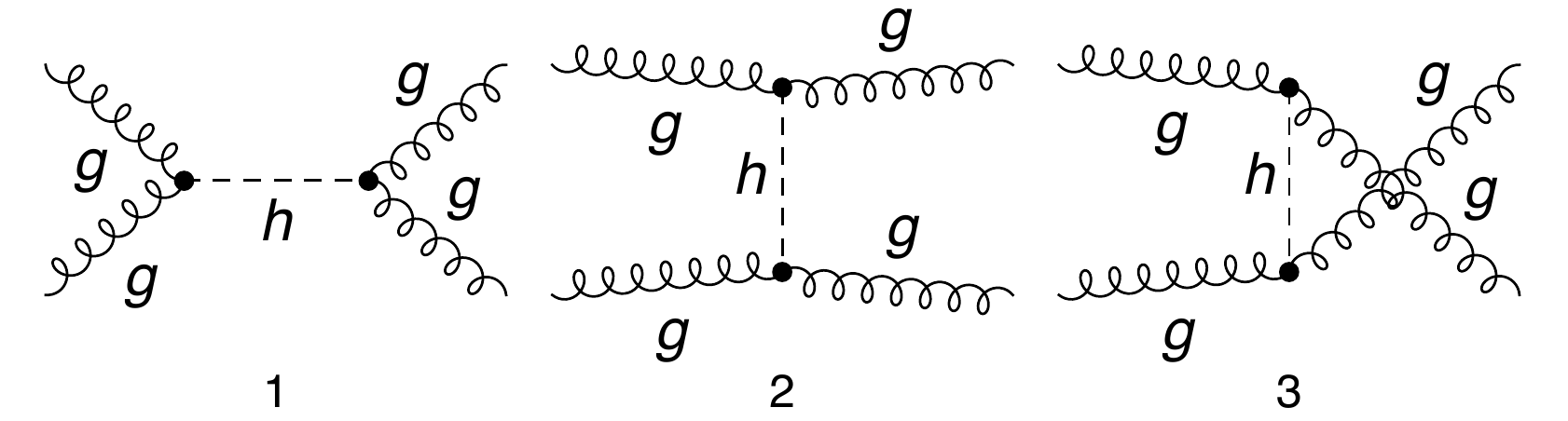}
	\caption{The agravity mediated $gg\rightarrow gg$ scattering. The curly and dashed lines represent the gluon and graviton propagators, respectively.}
	\label{fig01}
\end{figure}

The scattering amplitude for the process $gg \rightarrow gg$ mediated by a graviton in the $s$-channel, shown in Figure~\ref{fig01}.1, within the framework of agravity, can be expressed as
\begin{eqnarray}
\mathcal{M}_{s}&=&\frac{i}{12s^3} \left(f_0^2 \Pi^{\mu\nu\alpha\gamma}_1+f_2^2\Pi^{\mu\nu\alpha\gamma}_2\right)\epsilon_{a}^{\mu}(p_1)\epsilon_{b}^{\nu}(p_2)\epsilon_{c}^{*\alpha}(p_3)\epsilon_{d}^{*\gamma}(p_4) \delta^{a b} \delta^{c d}, 
\end{eqnarray}
where the tensors $\Pi^{\mu\nu\alpha\gamma}_1$ and $\Pi^{\mu\nu\alpha\gamma}_2$ are defined in the Appendix as functions of the external momenta $p_i$, and $s = (p_1 + p_2)^2 = (p_3 + p_4)^2$ is one of the Mandelstam variables. The diagrams corresponding to the other channels, shown in Figures~\ref{fig01}.2 and \ref{fig01}.3, are obtained from $\mathcal{M}_{s}$ via the following kinematical substitutions:
\begin{eqnarray}
\mathcal{M}_{t} & =& \mathcal{M}_{s}
\Big{|}_{
s \leftrightarrow t,~ 
p_3 \leftrightarrow -p_2,~b\leftrightarrow c
},\\
\mathcal{M}_{u} & =& \mathcal{M}_{s}
\Big{|}_{
u \leftrightarrow t,~ 
p_3 \leftrightarrow p_4,~c\leftrightarrow d
}.
\end{eqnarray}

Thus, the squared amplitude, defined as $|\mathcal{M}|^2 = |\mathcal{M}_s + \mathcal{M}_t + \mathcal{M}_u|^2$, can be expressed as
\begin{align}
\label{eq:new12}
|\mathcal{M}|^2 &= 
\frac{f_0^4\,t^2 u^2}{576\,s^4 t^2 u^2}
\Big[
3(979 N_c^2 - 675)t^4 
+ 4648 N_c^2 t^3 u 
+ 2(2336 N_c^2 + 2025)t^2 u^2 \nonumber\\
&\quad 
+ 4648 N_c^2 t u^3 
+ 3(979 N_c^2 - 675)u^4
\Big]
\nonumber\\
&\quad 
+ \frac{4 f_2^4}{576\,s^4 t^2 u^2}
\Big[ 
36 (N_c^2 - 1) t^8 
+ 180 (N_c^2 - 1) t^7 u 
+ 3(205 N_c^2 - 122) t^6 u^2 \nonumber\\
&\quad 
+ 4(244 N_c^2 - 45) t^5 u^3 
+ 2(581 N_c^2 + 18) t^4 u^4 
+ 4(244 N_c^2 - 45) t^3 u^5 \nonumber\\
&\quad 
+ 3(205 N_c^2 - 122) t^2 u^6 
+ 180 (N_c^2 - 1) t u^7 
+ 36 (N_c^2 - 1) u^8
\Big]
\nonumber\\
&\quad 
- \frac{8\,f_0^2 f_2^2\,t^2 u^2}{576\,s^4 t^2 u^2}
\Big[
3(109 N_c^2 - 67)t^4 
+ (511 N_c^2 - 36)t^3 u 
+ 2(209 N_c^2 + 315)t^2 u^2 \nonumber\\
&\quad 
+ (511 N_c^2 - 36)t u^3 
+ 3(109 N_c^2 - 67)u^4
\Big]\,,
\end{align}
\noindent
where $N_c$ denotes the number of colors, with $N_c = 3$ for QCD. The expression for $|\mathcal{M}|^2$ shown above corresponds to the squared matrix element fully summed over the color and polarization degrees of freedom of both the initial- and final-state gluons~\footnotemark.\footnotetext{\textit{Technical note on polarization sums.}
The original calculation replaced the polarization sum by \(\sum_\lambda \epsilon_\lambda^\mu(p)\epsilon_\lambda^{\nu *}(p) \to -\eta^{\mu\nu}\), which neglects the dependence on the gauge-fixing condition. The correct physical polarization sum reads
\[
\sum_\lambda \epsilon^\mu_\lambda(p) \epsilon^{\nu *}_\lambda(p)
= -\eta^{\mu\nu} + \frac{p^\mu n^\nu + p^\nu n^\mu}{p\cdot n}\,,
\]
where \(n^\mu\) is a reference vector satisfying \(n \cdot \epsilon(p) = 0\). We adopt the choice
\[
n_1^\mu = \frac{p_2^\mu}{\omega},\quad 
n_2^\mu =\frac{p_1^\mu}{\omega},\quad 
n_3^\mu =\frac{p_4^\mu}{\omega},\quad 
n_4^\mu =\frac{p_3^\mu}{\omega},
\]
valid in the CoM reference frame, where $\omega=\frac{\sqrt{s}}{2}$. This choice corresponds to the axial (physical) gauge and ensures that only the two transverse physical polarizations contribute to the scattering amplitude. It also eliminates spurious longitudinal components and leads to the correct tensor structure of the squared amplitude.}

It is important to note that, similarly to the ultra-Planckian scattering of photons~\cite{Holdom:2021oii}, the differential cross section for gluon scattering at ultra-Planckian energies scales as $1/E^2$ in the limit $E \rightarrow \infty$, as expected for a UV-complete theory of gravity. 

In order to investigate the conditions under which the squared amplitude $|\mathcal{M}|^2$ is positive, we evaluate it in the center-of-momentum (CoM) frame. In this frame, the Mandelstam variables are given by
\begin{eqnarray}
t = -\frac{s}{2}(1 - \cos{\theta}),\qquad u = -\frac{s}{2}(1 + \cos{\theta}).
\end{eqnarray}
Substituting these expressions into the amplitude, we obtain
\begin{align}
\label{ggggtheta}
|\mathcal{M}|^2
&= \frac{1}{12288\,\sin^4\theta}\Bigg[
49152\, f_2^4\!\left(5+3\cos 2\theta\right)
+\, \sin^4\theta\,\Big(
377529 f_0^4 - 334328 f_0^2 f_2^2 \nonumber\\
&\qquad + 158288 f_2^4
+ 20(7065 f_0^4 - 6632 f_0^2 f_2^2 + 1808 f_2^4)\cos 2\theta
\nonumber\\[1pt]
&\qquad
+\, 5(375 f_0^4 - 200 f_0^2 f_2^2 + 176 f_2^4)\cos 4\theta
\Big)\Bigg]\,,
\end{align}
\noindent where we have set $N_c = 3$, as required for QCD.

The expression \eqref{ggggtheta} is strictly positive for all scattering angles \(0<\theta<\pi\) and for arbitrary real couplings \(f_0\) and \(f_2\). The collinear structure is entirely contained in the prefactor of the first term, proportional to \(f_2^4\), leading to the behaviour \(|{\mathcal{M}}|^2 \sim f_2^4/\sin^4\theta\) near \(\theta \to 0,\pi\). This behaviour is the expected enhancement from spin-2 exchange in the \(t\)- and \(u\)-channels. In the limit where \(f_2\) becomes asymptotically small (as expected from asymptotic freedom), this enhancement is suppressed and the angular distribution remains finite for all physical scattering angles.



It is worth emphasizing that this latter scenario is quite plausible in the ultra-Planckian regime. As shown in Ref.~\cite{Salvio:2014soa}, the beta function for $f_0$ is positive, while the beta function for $f_2$ is negative. Consequently, $f_0$ increases with energy scale, whereas $f_2$ exhibits asymptotic freedom.

In addition, the authors of Ref.~\cite{Falls:2018ylp} investigate the asymptotic safety conjecture for quantum gravity by analyzing polynomial actions in the Ricci scalar. Their results provide strong evidence for the existence of a nontrivial UV fixed point, thereby supporting the viability of asymptotically safe quantum gravity even in the presence of high-order curvature invariants. This result suggests that the coupling $f_0$ may remain bounded and not diverge as the energy scale $E$ approaches infinity.

In the forward limit $t \to 0$ and under the assumption $f_0 \gg f_2$, the squared amplitude behaves as
\begin{align}
\label{eq:new15}
|\mathcal{M}|^2 &= f_2^4 (N_c^2 - 1)\left( \frac{s^2}{4 t^2} + \frac{s}{4 t} \right)\nonumber\\
&
+\frac{1}{192}\Big[ 
f_0^4 \big(979\,N_c^2 - 675\big) 
+ 4 f_2^4 \big(85\,N_c^2 - 2\big)
- 8 f_0^2 f_2^2 \big(109\,N_c^2 - 67\big) 
\Big]+\mathcal{O}(t/s).
\end{align}
\noindent The divergent $1/t$ term originates from the exchange of a massless graviton in the $t$-channel, as in GR, and represents the expected infrared behavior due to long-range gravitational interactions. The constant term, by contrast, reflects the contributions of the massive spin-2 and spin-0 states present in agravity\footnotemark.\footnotetext{In the formulation of agravity~\cite{Salvio:2014soa,Salvio:2017qkx}, the Einstein--Hilbert term is absent, and no massive spin-2 or spin-0 poles are present in the propagator. However, quantum effects can generate a mass scale dynamically via dimensional transmutation~\cite{Salvio:2020axm}, effectively restoring an Einstein term and leading to a spectrum that includes a massless graviton, a massive spin-2 ghost, and a scalar mode, as in quadratic gravity with explicit mass scales.}

The result \eqref{eq:new15} shows explicitly the emergence of collinear singularities as \(t \to 0\), controlled by the terms proportional to \(f_2^4\). If \(f_2\) is treated as a fixed, nonvanishing parameter, the amplitude diverges as \(t^{-2}\) and \(t^{-1}\), as expected from massless spin-2 exchange. However, in a realistic UV scenario where \(f_2\) is asymptotically free, these contributions become parametrically suppressed and the forward amplitude remains finite and IR safe. In this case, the scattering amplitude is well-behaved even for small scattering angles, consistent with the expected properties of a UV-complete dimensionless gravitational theory.

Following the approach proposed in Ref.~\cite{Holdom:2021oii}, we identify this constant piece as the effective forward scattering amplitude arising from the UV completion. This subtraction procedure allows us to isolate the genuinely new physics contributions from those already present in Einstein gravity. It is important to notice that the constant term is positive. After subtracting the IR-divergent term proportional to $1/t$ in the forward limit, the remaining contribution to the squared amplitude is independent of the CoM energy $s$. That is, $|\mathcal{M}(s, t \rightarrow 0)|^2 = \text{constant}$, implying that the forward amplitude $\mathcal{M}(s, 0)$ remains constant in the ultra-Planckian limit. Therefore, the Froissart bound is not violated in this case.

Finally, let us discuss some characteristics of the differential cross section. The expression for $|\mathcal{M}|^2$ presented above corresponds to the squared matrix element fully summed over the color and polarization degrees of freedom of both the initial- and final-state gluons. To obtain the unpolarized, color-averaged squared amplitude, it is necessary to divide this result by the total number of initial configurations. Since each incoming gluon carries two physical polarization states and $(N_c^2 - 1)$ color states, the total number of initial configurations is $4 (N_c^2 - 1)^2$. Therefore, the averaged squared amplitude is given by
\begin{equation}
|\mathcal{M}|^2_{\text{avg}} = \frac{1}{4 (N_c^2 - 1)^2} |\mathcal{M}|^2.
\end{equation}
This averaged result is the appropriate quantity to be used in the calculation of the differential cross section:
\begin{equation}\label{dsigmagggg}
\frac{d\sigma}{d\Omega} = \frac{1}{64\pi^2 s} \frac{1}{4 (N_c^2 - 1)^2} |\mathcal{M}|^2.
\end{equation}
\noindent In the particular case of QCD, where $N_c = 3$, one has $(N_c^2 - 1) = 8$, and the averaging factor becomes $\frac{1}{4 \times 64} = \frac{1}{256}$. 

It is straightforward to observe from Eqs.~\eqref{ggggtheta} and~\eqref{dsigmagggg} that the differential cross section scales as $1/s$ in the ultra-Planckian limit. This behavior is consistent with what is expected from a well behaved quantum theory of gravity. Notably, the result obtained for gluon scattering closely resembles that found in the case of photon-photon scattering, as discussed in Ref.~\cite{Holdom:2021oii}.

\subsection*{Gluon-Gluon Scattering into a Quark-Antiquark Pair}

\begin{figure}[h!]
	\includegraphics[angle=0 ,width=4.5cm]{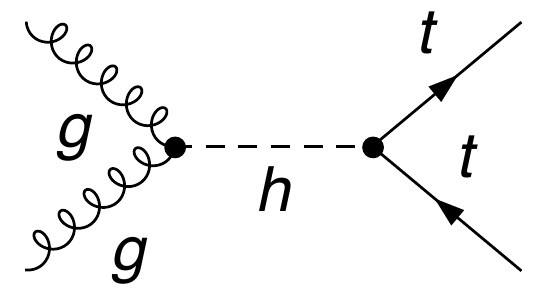}
	\caption{The agravity mediated $gg\rightarrow t\bar t$ scattering. The continuous lines represents the quark (antiquark) propagator.}
	\label{fig02}
\end{figure}

We now consider the process $gg \rightarrow q\bar{q}$ mediated by a graviton exchange. This channel provides a complementary probe of the gravitational interaction of matter fields at ultra-Planckian energies. The gluon-gluon scattering into a quark-antiquark pair is represented by the Feynman diagram shown in Figure~\ref{fig02}. The corresponding squared amplitude is given by
\begin{eqnarray}
|\mathcal{M}|^2 &=& \frac{f_2^4\, N_c\, \left(N_c^2 - 1\right) \left(t^2 + u^2\right) \left(s^2 - (t - u)^2\right)}{32 s^4},
\end{eqnarray}
\noindent where the expression implicitly includes the sum over the spin and color degrees of freedom of the final-state fermions, as well as the sum over the polarization and color states of the initial gluons. In order to obtain the unpolarized squared amplitude suitable for use in the cross section, one must divide by the number of initial configurations, given by $4(N_c^2 - 1)^2$.

In the CoM frame, this expression simplifies to
\begin{eqnarray}
|\mathcal{M}|^2 &=& \frac{f_2^4}{128} N_c \left(N_c^2 - 1\right) \sin^2{\theta} \left(\cos{2\theta} + 3\right).
\end{eqnarray}

It is straightforward to verify that $|\mathcal{M}|^2 \geq 0$ for all scattering angles in the range $0 \leq \theta \leq \pi$. Notably, the amplitude receives no contribution proportional to the coupling $f_0$, and depends solely on $f_2$. Since $f_2$ is asymptotically free, we find that $|\mathcal{M}|^2$ vanishes in the ultra-Planckian limit, i.e., $\lim_{E \rightarrow \infty} |\mathcal{M}|^2 = 0$.

Another important feature of this process is that the forward scattering limit, corresponding to $\theta \rightarrow 0$, is well behaved, as the squared amplitude vanishes in this limit: $\lim_{\theta \rightarrow 0} |\mathcal{M}|^2 = 0$. Furthermore, the differential cross section exhibits the same high-energy scaling behavior as in gluon-gluon scattering, falling off as $1/s$ in the ultra-Planckian regime.

\subsection*{Gluon-Quark Scattering}

We now analyze the process $gq \rightarrow gq$ mediated by a graviton exchange. This channel provides an additional test of the gravitational interaction of colored matter fields at ultra-Planckian energies, involving both gauge bosons and fermions in the initial and final states.

The squared amplitude for this process is given by
\begin{eqnarray}\label{m2gqgq}
|\mathcal{M}|^2 &=& \frac{f_2^4\, N_c\, \left(N_c^2 - 1\right) \left(s^2 + u^2\right) \left((s - u)^2 - t^2\right)}{32\, t^4},
\end{eqnarray}
\noindent
where the expression computed above includes the sum over the spin, polarization, and color degrees of freedom of both the initial- and final-state particles. To obtain the squared amplitude averaged over the final-state configurations, one must divide by the number of final degrees of freedom. For a final-state gluon and quark, this amounts to $4 N_c (N_c^2 - 1)$ configurations. The averaged squared amplitude is thus given by
\begin{equation}
|\mathcal{M}|^2_{\text{avg}} = \frac{1}{4 N_c (N_c^2 - 1)} |\mathcal{M}|^2.
\end{equation}

In the CoM frame, Eq.\eqref{m2gqgq} becomes
\begin{eqnarray}
|\mathcal{M}|^2 &=& \frac{f_2^4 N_c \left(N_c^2 - 1\right) (\cos{\theta} + 1) \left(\cos^2{\theta} + 2\cos{\theta} + 5\right)}{4 (\cos{\theta} - 1)^4}.
\end{eqnarray}

The above expression for $|\mathcal{M}|^2$ reveals a divergence in the forward scattering limit. In this regime, the denominator $(\cos{\theta} - 1)^4$ tends to zero, while the numerator remains finite, leading to
$ |\mathcal{M}|^2 \sim \frac{1}{(\cos{\theta} - 1)^4} \rightarrow \infty \quad \text{as} \quad \theta \rightarrow 0$.  

When the quarks are taken to be massive, the forward scattering behavior of the graviton-mediated quark–gluon process is qualitatively modified. In the massless case, the squared amplitude diverges in the forward limit due to the vanishing of the Mandelstam variable $t$ as $\theta \rightarrow 0$, resulting in an IR singularity. However, the presence of a nonzero quark mass $m_q$ alters the kinematics such that $t$ no longer tends to zero in this limit. Specifically, in the CoM frame, the Mandelstam variable is given by $t = m_q^2 - 2 E^2 (1 - \beta \cos\theta)$, with $\beta = \sqrt{1 - \frac{m_q^2}{E^2}} < 1$, which ensures that $t$ remains finite as $\theta \rightarrow 0$. Consequently, the squared amplitude $|\mathcal{M}|^2$ is regularized, and the differential cross section remains finite at all scattering angles. The quark mass thus acts as a natural IR regulator, suppressing the forward divergence and rendering the process physically well-defined. Moreover, in the ultra-Planckian limit, the differential cross section continues to scale as $1/s$, consistent with expectations for a UV-complete theory of gravity.

\subsection*{Quark-Quark Scattering into Two Quarks}

We now turn to the process $qq \rightarrow qq$, mediated by a graviton exchange. This process offers further insight into the gravitational interactions among fermionic matter fields at ultra-Planckian energies.

The squared amplitude for quark–quark scattering via graviton exchange is given by
\begin{eqnarray}\label{qqqq_scat}
|\mathcal{M}|^2 &=& \frac{f_2^4\, N_c}{64\, t^4 u^4} \Big[
N_c\, t^4 u^4 + 16 N_c\, s^4 \left(t^4 + u^4\right) + 32 N_c\, s^3 t u \left(t^3 + u^3\right) \nonumber\\
&& + (21 N_c + 4) s^2 t^2 u^2 \left(t^2 + u^2\right) - (5 N_c - 17) s^2 t^3 u^3
\Big],
\end{eqnarray}
\noindent
where the expression corresponds to the squared matrix element summed over the spin and color degrees of freedom of both the initial- and final-state quarks. In order to obtain the squared amplitude averaged over the final-state configurations, one must divide by the number of final degrees of freedom. For two final-state quarks, this amounts to $4 N_c^2$ configurations. Therefore, the averaged squared amplitude is given by
\begin{equation}
|\mathcal{M}|^2_{\text{avg}} = \frac{1}{4 N_c^2} |\mathcal{M}|^2.
\end{equation}

Specializing to QCD with $N_c = 3$ and using the Mandelstam identity $s = -t - u$, the amplitude simplifies to
\begin{eqnarray}
|\mathcal{M}|^2 &=& \frac{3 f_2^4}{64\, t^4 u^4} \Big[
48 t^8 + 96 t^7 u + 67 t^6 u^2 + 40 t^5 u^3 + 45 t^4 u^4 \nonumber\\
&& + 40 t^3 u^5 + 67 t^2 u^6 + 96 t u^7 + 48 u^8
\Big],
\end{eqnarray}
\noindent
where the symmetry under the exchange $t \leftrightarrow u$ is manifest, as expected for the scattering of identical fermions in both initial and final states.

In the CoM frame, the squared amplitude Eq.\eqref{qqqq_scat} takes the form
\begin{eqnarray}
|\mathcal{M}|^2 &=& \frac{f_2^4 N_c}{64 \sin^8{\theta}} \Big[
N_c \cos^8{\theta} + 4(46 N_c - 9) \cos^6{\theta} + 2(913 N_c + 86) \cos^4{\theta} \nonumber\\
&& + 4(484 N_c - 59) \cos^2{\theta} + 149 N_c + 100
\Big].
\end{eqnarray}

It is important to note that the results discussed above were obtained in the massless quark limit. In this regime, the squared amplitude exhibits an infrared divergence in the forward direction, as $t \rightarrow 0$ when $\theta \rightarrow 0$. However, the inclusion of nonzero quark masses modifies the kinematics of the scattering process, introducing a natural infrared cutoff as discussed before. In particular, the Mandelstam variable $t$ no longer vanishes in the forward limit due to the finite mass, and as a consequence, the divergence in $|\mathcal{M}|^2$ is regulated. This leads to a finite and well behaved differential cross section for all scattering angles. The mass of the quark thus plays the role of an infrared regulator, softening the singular behavior of the gravitational interaction in the $t$-channel and rendering the forward scattering physically meaningful.

\section{Final Remarks}\label{summary}

In this work, we have investigated several scattering processes involving quarks and gluons mediated by graviton exchange within the framework of agravity—a scale-invariant and renormalizable theory of quadratic quantum gravity. Focusing on the ultra-Planckian regime, we computed the tree-level amplitudes for $gg \rightarrow gg$, $gg \rightarrow q\bar{q}$, $gq \rightarrow gq$, and $qq \rightarrow qq$, and analyzed the behavior of the squared amplitudes and the corresponding differential cross sections.

One of the questions addressed in this paper concerns the compatibility between unitarity and UV completeness in quadratic gravity models that inherently include ghostlike states. By isolating the gravitational contributions and examining the positivity properties of the squared amplitudes, we assessed whether unitarity-preserving behavior can persist in spite of the presence of higher-derivative ghosts. Our results indicate that the squared amplitudes remain positive in a wide region of the angular domain, particularly when the gravitational couplings satisfy $f_2 \ll f_0$, as is naturally expected due to their renormalization group running in the deep UV. The analysis also shows that the differential cross sections for all studied processes scale as $1/s$ in the high-energy limit. This behavior, consistent with a Froissart-like bound, supports the interpretation of agravity as a UV-complete framework.

An important observation concerns the forward scattering behavior. For processes involving massless particles, such as $gg \rightarrow gg$ and $gq \rightarrow gq$, we identified divergences in the forward limit, which are typical IR signatures of long-range interactions mediated by massless gravitons. However, the presence of quark masses for $gq \rightarrow gq$ process provides a natural IR regulator, taming these divergences and ensuring a physically meaningful cross section at all angles. Furthermore, by isolating and subtracting the IR-divergent $1/t$ term in the forward amplitude between two gluons, $gg \rightarrow gg$, we were able to extract a finite, constant contribution that reflects the genuine UV effects of the higher-derivative gravitational modes. Our findings confirm and extend previous results obtained in photon-photon scattering~\cite{Holdom:2021oii,Holdom:2021hlo}, demonstrating that unitarity can be preserved in quadratic gravity even in the absence of manifest positivity. The fact that similar cancellations and scaling behaviors persist in the context of non-Abelian gauge interactions supports the universality of these mechanisms and reinforces the physical viability of agravity at trans-Planckian energies.

We emphasize that the amplitudes computed here are purely gravitational, with no interference from gauge-mediated contributions. This choice allowed us to cleanly assess the effects of the graviton sector alone. Future directions include exploring interference effects, loop corrections, and the impact of spontaneous symmetry breaking or conformal anomalies, all of which could enrich the structure of gravitationally induced interactions in quantum field theory.

Overall, our analysis provides additional evidence that quadratic gravity, and agravity in particular, may offer a self-consistent, UV-complete, and perturbatively unitary extension of gravity compatible with high-energy particle physics.

\acknowledgments
I.F.C. is partially supported by Coordena\c{c}\~ao de Aperfei\c{c}oamento de Pessoal de N\'ivel Superior (CAPES). The work of A. C. L. has been partially supported by the CNPq project No. 404310/2023-0. 

\appendix

\section*{Appendix: Tensor Structures in the $gg \rightarrow gg$ Scattering Amplitude}

In this appendix, we present the explicit forms of the tensor structures $\Pi^{\mu\nu\alpha\gamma}_1$ and $\Pi^{\mu\nu\alpha\gamma}_2$ that appear in the $s$-channel amplitude for gluon–gluon scattering mediated by a graviton within the framework of agravity. These tensors arise from the contraction between the graviton propagator and the energy-momentum tensors of the external gluons. The expressions are written in terms of the external momenta and the Mandelstam variables, defined as
\begin{equation}
s = (p_1 + p_2)^2 = (p_3 + p_4)^2, \quad
t = (p_1 - p_3)^2 = (p_4 - p_2)^2, \quad
u = (p_1 - p_4)^2 = (p_3 - p_2)^2.
\end{equation}

The tensor $\Pi^{\mu\nu\alpha\gamma}_1$ is given by:
\begin{eqnarray}
\Pi^{\mu\nu\alpha\gamma}_1&=& - 
\left( 
5s\, \eta^{\alpha\gamma} 
- 2\,  p_4^\alpha p_3^\gamma 
- 18 \, p_3^\alpha p_4^\gamma 
\right)
\left(
5 s^2 \, \eta^{\mu\nu}
- 20s \, p_2^\mu p_1^\nu 
- 2t \, p_3^\mu p_1^\nu 
 \right.\nonumber\\
&&\left. - 2u \, p_3^\mu p_1^\nu
- 2t \, p_4^\mu p_1^\nu 
- 2u \, p_4^\mu p_1^\nu 
- 2t \, p_2^\mu p_3^\nu 
- 2u \, p_2^\mu p_3^\nu 
- 2s \, p_3^\mu p_3^\nu 
 \right.\nonumber\\
&& \left.
- 2s \, p_4^\mu p_3^\nu
- 2t \, p_2^\mu p_4^\nu
- 2u \, p_2^\mu p_4^\nu 
- 2s \, p_3^\mu p_4^\nu 
- 2s \, p_4^\mu p_4^\nu
\right).
\end{eqnarray}

The tensor $\Pi^{\mu\nu\alpha\gamma}_2$ contains a more elaborate structure. Its full expression is given by
\begin{eqnarray}
\Pi^{\mu\nu\alpha\gamma}_2&=&  s^3\left( 3 \, \eta^{\alpha\nu} \eta^{\gamma\mu} + 3 \, \eta^{\alpha\mu} \eta^{\gamma\nu} - 2 \, \eta^{\alpha\gamma} \eta^{\mu\nu} \right)
+s^2
\Big( 2 \, p_4^\alpha p_3^\gamma \eta^{\mu\nu} + 6 \, p_3^\alpha p_4^\gamma \eta^{\mu\nu} 
- 6 \, p_1^\alpha \eta^{\gamma\nu} p_2^\mu \nonumber\\
&&- 6 \, \eta^{\alpha\nu} p_1^\gamma p_2^\mu 
- 3 \, p_4^\alpha \eta^{\gamma\nu} p_3^\mu 
- 3 \, \eta^{\alpha\nu} p_3^\gamma p_3^\mu 
- 6 \, \eta^{\alpha\nu} p_4^\gamma p_3^\mu 
- 6 \, p_3^\alpha \eta^{\gamma\nu} p_4^\mu 
- 3 \, p_4^\alpha \eta^{\gamma\nu} p_4^\mu  \nonumber\\
&& - 3 \, \eta^{\alpha\nu} p_3^\gamma p_4^\mu 
- 6 \, p_2^\alpha \eta^{\gamma\mu} p_1^\nu 
- 6 \, \eta^{\alpha\mu} p_2^\gamma p_1^\nu 
+ 8 \, \eta^{\alpha\gamma} p_2^\mu p_1^\nu 
- 3 \, p_4^\alpha \eta^{\gamma\mu} p_3^\nu 
- 3 \, \eta^{\alpha\mu} p_3^\gamma p_3^\nu 
\nonumber\\
&&- 6 \, \eta^{\alpha\mu} p_4^\gamma p_3^\nu 
+ 2 \, \eta^{\alpha\gamma} p_3^\mu p_3^\nu 
+ 2 \, \eta^{\alpha\gamma} p_4^\mu p_3^\nu 
- 6 \, p_3^\alpha \eta^{\gamma\mu} p_4^\nu 
- 3 \, p_4^\alpha \eta^{\gamma\mu} p_4^\nu 
- 3 \, \eta^{\alpha\mu} p_3^\gamma p_4^\nu \nonumber\\
&& + 2 \, \eta^{\alpha\gamma} p_3^\mu p_4^\nu + 2 \, \eta^{\alpha\gamma} p_4^\mu p_4^\nu \Big)
-s 
 \Big( t \big( 3\, p_4^\alpha \eta^{\gamma\nu} p_2^\mu 
 + 3\, \eta^{\alpha\nu} p_3^\gamma p_2^\mu 
 + 6\, \eta^{\alpha\nu} p_4^\gamma p_2^\mu 
 \nonumber\\
 &&+ 6\, p_3^\alpha \eta^{\gamma\mu} p_1^\nu + 3\, p_4^\alpha \eta^{\gamma\mu} p_1^\nu + 3\, \eta^{\alpha\mu} p_3^\gamma p_1^\nu - 2\, \eta^{\alpha\gamma} p_3^\mu p_1^\nu - 2\, \eta^{\alpha\gamma} p_4^\mu p_1^\nu - 2\, \eta^{\alpha\gamma} p_2^\mu p_3^\nu \nonumber\\
 && - 2\, \eta^{\alpha\gamma} p_2^\mu p_4^\nu \big) 
 + u \big( 6\, p_3^\alpha \eta^{\gamma\nu} p_2^\mu 
 + 3\, p_4^\alpha \eta^{\gamma\nu} p_2^\mu 
 + 3\, \eta^{\alpha\nu} p_3^\gamma p_2^\mu 
 + 3\, p_4^\alpha \eta^{\gamma\mu} p_1^\nu 
 + 3\, \eta^{\alpha\mu} p_3^\gamma p_1^\nu \nonumber\\
 && + 6\, \eta^{\alpha\mu} p_4^\gamma p_1^\nu - 2\, \eta^{\alpha\gamma} p_3^\mu p_1^\nu - 2\, \eta^{\alpha\gamma} p_4^\mu p_1^\nu - 2\, \eta^{\alpha\gamma} p_2^\mu p_3^\nu - 2\, \eta^{\alpha\gamma} p_2^\mu p_4^\nu \big) 
 + \big( 8\, p_4^\alpha p_3^\gamma p_2^\mu p_1^\nu \nonumber\\
 &&+ 24\, p_3^\alpha p_4^\gamma p_2^\mu p_1^\nu - 6\, p_4^\alpha p_2^\gamma p_3^\mu p_1^\nu - 6\, p_2^\alpha p_3^\gamma p_3^\mu p_1^\nu - 12\, p_2^\alpha p_4^\gamma p_3^\mu p_1^\nu - 12\, p_3^\alpha p_2^\gamma p_4^\mu p_1^\nu \nonumber\\
 &&- 6\, p_4^\alpha p_2^\gamma p_4^\mu p_1^\nu - 6\, p_2^\alpha p_3^\gamma p_4^\mu p_1^\nu - 6\, p_4^\alpha p_1^\gamma p_2^\mu p_3^\nu - 6\, p_1^\alpha p_3^\gamma p_2^\mu p_3^\nu - 12\, p_1^\alpha p_4^\gamma p_2^\mu p_3^\nu \nonumber\\
 &&- 4\, p_4^\alpha p_3^\gamma p_3^\mu p_3^\nu - 12\, p_4^\alpha p_4^\gamma p_3^\mu p_3^\nu - 6\, p_3^\alpha p_3^\gamma p_4^\mu p_3^\nu - 4\, p_4^\alpha p_3^\gamma p_4^\mu p_3^\nu - 6\, p_4^\alpha p_4^\gamma p_4^\mu p_3^\nu \nonumber\\
 &&- 12\, p_3^\alpha p_1^\gamma p_2^\mu p_4^\nu - 6\, p_4^\alpha p_1^\gamma p_2^\mu p_4^\nu - 6\, p_1^\alpha p_3^\gamma p_2^\mu p_4^\nu - 6\, p_3^\alpha p_3^\gamma p_3^\mu p_4^\nu - 4\, p_4^\alpha p_3^\gamma p_3^\mu p_4^\nu \nonumber\\
 && - 6\, p_4^\alpha p_4^\gamma p_3^\mu p_4^\nu - 12\, p_3^\alpha p_3^\gamma p_4^\mu p_4^\nu - 4\, p_4^\alpha p_3^\gamma p_4^\mu p_4^\nu \big) \Big)
 +2 \Big(
  u \big( 2\, p_4^\alpha p_3^\gamma p_3^\mu p_1^\nu + 6\, p_4^\alpha p_4^\gamma p_3^\mu p_1^\nu 
  \nonumber\\
  && + 3\, p_3^\alpha p_3^\gamma p_4^\mu p_1^\nu 
  + 2\, p_4^\alpha p_3^\gamma p_4^\mu p_1^\nu 
  + 3\, p_4^\alpha p_4^\gamma p_4^\mu p_1^\nu 
  + 3\, p_3^\alpha p_3^\gamma p_2^\mu p_3^\nu 
  + 2\, p_4^\alpha p_3^\gamma p_2^\mu p_3^\nu 
  \nonumber\\
  && + 3\, p_4^\alpha p_4^\gamma p_2^\mu p_3^\nu + 6\, p_3^\alpha p_3^\gamma p_2^\mu p_4^\nu + 2\, p_4^\alpha p_3^\gamma p_2^\mu p_4^\nu \big) 
  + t \big( 3\, p_3^\alpha p_3^\gamma p_3^\mu p_1^\nu + 2\, p_4^\alpha p_3^\gamma p_3^\mu p_1^\nu 
  \nonumber\\
  && + 3\, p_4^\alpha p_4^\gamma p_3^\mu p_1^\nu 
  + 6\, p_3^\alpha p_3^\gamma p_4^\mu p_1^\nu 
  + 2\, p_4^\alpha p_3^\gamma p_4^\mu p_1^\nu 
  + 2\, p_4^\alpha p_3^\gamma p_2^\mu p_3^\nu 
  + 6\, p_4^\alpha p_4^\gamma p_2^\mu p_3^\nu 
  \nonumber\\
  &&+ 3\, p_3^\alpha p_3^\gamma p_2^\mu p_4^\nu + 2\, p_4^\alpha p_3^\gamma p_2^\mu p_4^\nu + 3\, p_4^\alpha p_4^\gamma p_2^\mu p_4^\nu \big) 
 \Big).
\end{eqnarray}
\noindent
These expressions were generated using algebraic packages~\cite{feyncalc,Hahn:1998yk,feyncalc1,feyncalc2,feynarts,feynrules,feynrules1,feynhelpers} and enter directly in the computation of the amplitude for the $gg \rightarrow gg$ process.

Note that $\Pi^{\mu\nu\alpha\gamma}_1$ and $\Pi^{\mu\nu\alpha\gamma}_2$ encode the part of kinematical dependence of the external gluons. Their contractions with polarization vectors yield the graviton-mediated amplitude in the $s$-channel, from which the $t$- and $u$-channel contributions are obtained by crossing symmetry.

\newpage

\begin{figure}[h!]
	\includegraphics[angle=0 ,width=4.5cm]{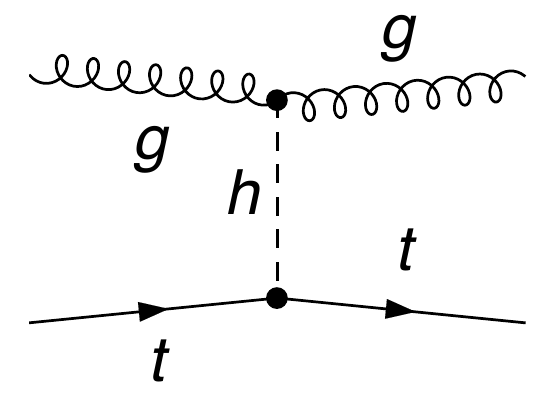}
	\caption{The agravity mediated $gt\rightarrow gt$ scattering.}
	\label{fig03}
\end{figure}

\begin{figure}[h!]
	\includegraphics[angle=0 ,width=9cm]{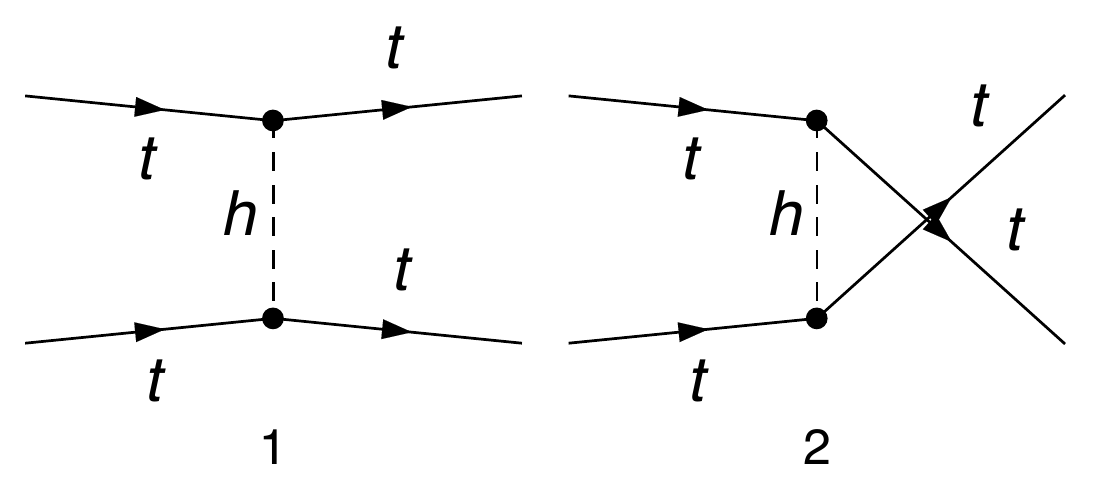}
	\caption{The agravity mediated $tt\rightarrow tt$ scattering.}
	\label{fig04}
\end{figure}

\end{document}